\documentclass[a4paper,twocolumn,amsmath,amssymb,superscriptaddress,longbibliography,pra, nofootinbib]{revtex4-1}
\usepackage{bm}% bold math
\usepackage{braket}
\usepackage{dsfont}
\usepackage[usenames,dvipsnames]{xcolor}
\usepackage{pstricks}
\usepackage[tight]{subfigure}
\usepackage{verbatim}
\usepackage{units}
\usepackage{natbib}
\usepackage{multirow}
\usepackage{enumitem}
\usepackage{mathrsfs}
\usepackage{wasysym}
\usepackage{leftidx}
\usepackage{xspace}
\usepackage{graphicx}

\usepackage[normalem]{ulem}  % gives command \sout = strike out text

\usepackage[a4paper]{hyperref}
\hypersetup{colorlinks=true,linktoc=all,linkcolor=blue,breaklinks=true,citecolor=blue,urlcolor=blue}
\usepackage{siunitx}

\AtBeginDocument{%
    \newwrite\bibnotes
    \def\bibnotesext{Notes.bib}
    \immediate\openout\bibnotes=\jobname\bibnotesext
    \immediate\write\bibnotes{@CONTROL{REVTEX41Control}}
    \immediate\write\bibnotes{@CONTROL{%
    apsrev41Control,author="08",editor="1",pages="1",title="0",year="1"}}
     \if@filesw
     \immediate\write\@auxout{\string\citation{apsrev41Control}}%
    \fi
}%

\newcommand{\bea}{\begin{eqnarray}}
\newcommand{\eea}{\end{eqnarray}}
\newcommand{\be}{\begin{equation}}
\newcommand{\ee}{\end{equation}}
\newcommand{\benn}{\begin{equation*}}
\newcommand{\eenn}{\end{equation*}}

\renewcommand{\Re}{\mathop{\mathrm{Re}}}

\newcommand{\kBT}{k_\text{B}T}

\definecolor{light-gray}{gray}{0.9}
\definecolor{gris}{gray}{0.5}
\definecolor{DarkGreen}{rgb}{0.,0.4,0.4}
\definecolor{amber}{rgb}{1,0.75,0}
\definecolor{brown}{rgb}{0.65, 0.16, 0.16}

\begin{document}

\title{%Andreev-Coulomb heat engine\\
%Superconducting hybrid quantum dot heat engine\\
Nonlocal quantum heat engines with hybrid superconducting devices}
\author{S. Mojtaba Tabatabaei}
\affiliation{Department of Physics, Kharazmi University, 15719-14911 Tehran, Iran}
\author{David S\'anchez}
\affiliation{Institute for Cross-Disciplinary Physics and Complex Systems IFISC (UIB-CSIC), E-07122 Palma de Mallorca, Spain}
\author{Alfredo Levy Yeyati}
\affiliation{Departamento de F\'isica Te\'orica de la Materia Condensada, Condensed Matter Physics Center (IFIMAC), and Instituto Nicol\'as Cabrera, Universidad Aut\'onoma de Madrid, 28049 Madrid, Spain\looseness=-1}
\author{Rafael S\'anchez}
\affiliation{Departamento de F\'isica Te\'orica de la Materia Condensada, Condensed Matter Physics Center (IFIMAC), and Instituto Nicol\'as Cabrera, Universidad Aut\'onoma de Madrid, 28049 Madrid, Spain\looseness=-1}
\date{\today}

\begin{abstract}
We discuss a quantum thermal machine that generates power
from a thermally driven double quantum dot coupled to
normal and superconducting reservoirs.
Energy exchange between the dots is mediated by
electron-electron interactions. We can distinguish
three main mechanisms within the device operation modes.
In the Andreev tunneling regime, energy flows
in the presence of coherent superposition of zero-
and two-particle states. Despite the intrinsic
electron-hole symmetry of Andreev processes,
we find that the heat engine efficiency increases
with increasing coupling to the superconducting reservoir.
The second mechanism occurs in the regime of quasiparticle transport. Here we obtain large efficiencies due to the presence of the superconducting gap and the strong
energy dependence of the electronic density of states around the gap edges. Finally, in the third regime there exists a competition between Andreev processes and quasiparticle tunneling. 
%\sout{While the first two regimes are correctly described with a master equation approach, the intermediate situation requires the application of a nonequilibrium Green's function technique with dynamical self-energies.} 
Altogether, our results emphasize the importance
of both pair tunneling and structured band spectrum for an accurate characterization of the heat engine properties in normal-superconducting coupled dot systems.
\end{abstract}
\maketitle

\section{Introduction}
\label{sec:intro}

Thermoelectric effects in solid state devices allow to convert heat exchanged with the environment into an electric current.
Nanoscale conductors can this way work as on-chip converters of waste heat into power at low temperatures~\cite{benenti:2017}. To this end, a mechanism that breaks particle-hole symmetry is needed. Among the desired properties of a good thermoelectric engine, the conductor should be electrically isolated from the thermal source, such that the absorbed heat and the generated charge currents are well separated as is the case, e.g., in a thermocouple. Despite the fact that superconductors are good thermal insulators and (obviously) good electrical conductors~\cite{mazza:2015}, they are rarely considered as components of thermoelectric generators due to their intrinsic electron-hole symmetry~\cite{ginzburg1991thermoelectric}. 

Recent approaches to this problem exploit tunnel junctions between unequal superconducting electrodes using nonlinearities~\cite{marchegiani_nonlinear_2020,marchegiani_phase_2020,germanese_bipolar_2022} or single-electron transistors~\cite{donald,kamp_phase_2019,bauer_phase_2021}, hybrid normal-superconductor junctions including quantum dot energy filters~\cite{Cao2015,sanchez_cooling_2018,Hussein2019,Kirsanov2019,Tan2021,verma:2022}, spin-dependent scattering~\cite{machon_nonlocal_2013,giazotto_proposal_2014,ozaeta:2014,hwang:2016,hwang:2016b,kolenda:2016,marchegiani:2016,savander_thermoelectric_2020,keidel_ondemand_2020,Blasi2020a,blasi_nonlocal_2021}, photon-assisted tunneling~\cite{hofer:2016prb}, or interference phenomena~\cite{jacquod_coherent_2010,kalenkov_large_2017}. The gapped density of states has also been successfully used for cooling~\cite{Nahum1994Dec,Leivo1996Apr} and for power sources~\cite{Marin-Suarez:2022,pekola:2022}.
%,timofeev_electronic_2009}

Mesoscopic versions of the thermocouple geometry are based on three terminal configurations, where one terminal acts as the thermal reservoir and the other two support the charge current. The coupling to the heat source must be such that no particle currents are injected in the conductor, at least on average. Different mechanisms have been proposed depending on the way carriers couple to the heat source. This coupling can be mediated by microscopic interactions, e.g, charge correlations~\cite{hotspots,sothmann:2012,thierschmann:2015,hartmann:2015,roche:2015,thierschmann_thermal_2015} and electron-boson interactions~\cite{rutten:2009,entin:2010,krause:2011,sothmann:2012epl,ruokola:2012,jiang:2012,bergenfeldt:2014,arrachea:2014,bosisio_nanowire_2016,yamamoto:2016,jiang:2017,dorsch:2020,dorsch_characterization_2021}, or due to electronic relaxation in hot probes~\cite{jordan:2013,mazza:2014,donsa:2014,jiang_enhancing_2014,jaliel:2019}, %\textcolor{blue}{what is the difference between bosonic excitations and inelastic scattering?}
%or via local injectors~\cite{harzheim:2018,fast:2020,genevieveqpc,extrinsic}, 
to mention a few quantum dot systems.
Of particular interest are heat engines based on capacitively coupled  dots~\cite{molenkamp:1995,chan:2002,hubel:2007,mcclure:2007}, where the separation of charge and heat currents is explicit: current flows in one conductor, that we call passive, with the other one being connected to the heat source via charge fluctuations~\cite{hotspots,sothmann:2012,thierschmann:2015,hartmann:2015,roche:2015,strasberg:2013,zhang:2016,dare:2017,walldorf:2017,strasberg:2018,mayrhofer:2021}. Reversely the charge current (in response to a voltage bias or to temperature differences) can be used to operate the system as a refrigerator~\cite{sanchez:2013,zhang:2015,koski:2015,sanchez_correlation_2017,erdman_absorption_2018,dare:2019}.
%,sanchezd:2019}. 

\begin{figure}[b]
\includegraphics[width=\linewidth]{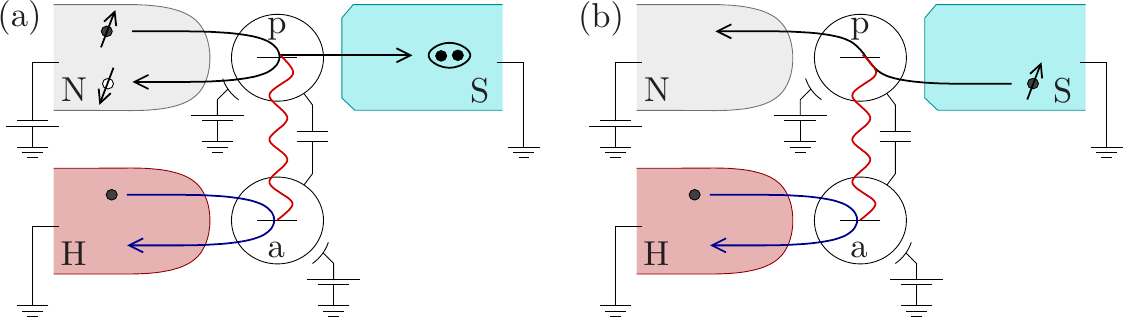}
\caption{\label{fig:scheme}Hybrid normal-superconductor heat engine based on coupled quantum dots ($p$, for passive and $a$, for active). Heat injected from the hot reservoir (H) induces the transfer of either (a) a Cooper pair in the superconductor (S) via an Andreev reflection in the normal lead (N), or (b) a quasiparticle transferred between S and N. In both cases, an electric current is generated in the passive subsystem.
} 
\end{figure}

Similar coupling schemes are relevant for mesoscopic Coulomb drag configurations~\cite{mortensen_coulomb_2001,khrapai:2006,gasser:2009,shinkai:2009,Moldoveanu2009Jul,sanchez:2010,hussein:2012,hussein:2015,bischoff:2015,kaasbjerg:2016,keller:2016,lim:2018,sierra:2019,takada:2021} which generate a current in the passive circuit by coupling it to another one (the drive) that is voltage biased. Differently from extended samples, that rely on momentum exchange~\cite{Narozhny2016May}, the mesoscopic drag is based on the exchange of energy. This has been emphasized in proposals of thermal drag currents where only heat flows in the drive system~\cite{whitney:2015,lu:2016,bhandari:2018,ben-abdallah:2019,sanchez:2019,berdanier:2019,Lu:2021,Xi:2021,wang:2022,idrisov:2022} and of absorption refrigerators~\cite{venturelli:2013}. %\textcolor{red}{Several experiments have explored this effect also in extended samples~\cite{...,Tao_josephson_2020}.?} 
In a recent work~\cite{mojtaba}, we found that the interplay of charge fluctuations and Andreev reflection processes gives rise to a drag current when the passive system contains a superconducting electrode. These processes compete with a second mechanism due to single quasiparticle tunneling. Here we explore how these mechanisms can make the system work as a nonlocal thermoelectric heat engine. For this purpose, we consider a hybrid three-terminal configuration consisting of two Coulomb-coupled quantum dots ($a$, for active and $p$, for passive), as sketched in Fig.~\ref{fig:scheme}. The active dot is coupled to a hot normal terminal (H). The passive one is connected to one normal (N) and one superconducting (S) terminals, see Refs.~\cite{graber:2004,deacon:2010,deFranceschi:2010} for discussions of related experimental implementations. We analyze the regimes where the generated current is mediated by the two mentioned mechanisms [Andreev reflection, cf. Fig.~\ref{fig:scheme}(a), and correlated quasiparticles filtered by the superconducting density of states, cf. Fig.~\ref{fig:scheme}(b)]. They are shown to give opposite contributions, which  in the intermediate regime where the two processes coexist reduces the thermoelectric performance of the engine (in terms of the generated power and efficiency). 

Additionally it is interesting to explore the properties of the heat currents, which may make the device work as a refrigerator (when heat is extracted from the coldest reservoir) or as a heat pump (when heat flows into the hottest one). Multitask operations in three terminal conductors have been recently identified~\cite{manzano:2020} where combinations of two or more such operations are met, see also Refs.~\cite{entin:2015,hammam_2022}. 

The remaining of the manuscript is organized as follows. The system and the model are described in Sec.~\ref{sec:model}. The limiting regimes where Andreev and quasiparticle transport dominate are discussed in Sec.~\ref{sec:andreev} and \ref{sec:quasipart} respectively, by using a master equation approach~\cite{Schaller2014,Strasberg2022} that provides a physical understanding of the involved mechanisms. The two regimes require different master equations which will be discussed separately. The intermediate regime is explored numerically in Sec.~\ref{sec:keldysh} by invoking a non-equilibrium Green's functions technique~\cite{Haug2008}. Finally, conclusions are presented in Sec.~\ref{sec:conclusions}.

%\textcolor{red}{Josephson-Coulomb drag~\cite{Tao_josephson_2020}}

\section{Hybrid coupled quantum dot system}
\label{sec:model}

\subsection{Hamiltonian}

We aim at the simplest description of the main mechanisms involved in the passive current generation. Spin is essential for pairing in the passive system. In the active dot, only charge fluctuations are required in order to generate a passive current~\cite{sierra:2019}.  Therefore, we ignore the spin degree of freedom and intradot Coulomb interactions in the active system for simplicity. This provides a good description of transport (up to spin degeneracy prefactors) as long as double occupancy of the active dot is negligible and in the absence of interdot spin-spin interactions. Note however that intradot Coulomb interactions in the passive dot do affect the pairing processes and need to be taken into account.

Our system is hence described with the Hamiltonian
\begin{align}\label{eq_H}
{\cal H}={\cal H}_{l}
+{\cal H}_{{\rm {dqd}}}+{\cal H}_{t}.    
\end{align}
The first term is the Hamiltonian for the reservoirs,
\begin{gather}
\begin{aligned}
{\cal H}_{l}&= \sum_{k,\sigma}\varepsilon_{k,S}\hat{c}_{k,S,\sigma}^{\dagger}\hat{c}_{k,S,\sigma}+
\sum_k\Delta(\hat{c}_{k,S,\uparrow}^{\dagger}\hat{c}_{k,S,\downarrow}^{\dagger}+\rm{h.c.})\\
&+\sum_{k}\varepsilon_{k,H}\hat{c}_{k,H}^{\dagger}\hat{c}_{k,H}+
\sum_{k,\sigma}\varepsilon_{k,N}\hat{c}_{k,N,\sigma}^{\dagger}\hat{c}_{k,N,\sigma},
\label{eq:Hleads}
\end{aligned}
\end{gather}
where $\hat{c}_{k,\beta,\sigma}$ is the annihilation operator
for electrons with energy $\varepsilon_{k}$, momentum $k$
and spin $\sigma{=}\{\uparrow,\downarrow\}$ in terminal $\beta$=H,N,S (note that for $\beta$=H we drop the spin index),
and $\Delta$ is the order parameter in the superconducting electrode. Each of the metallic reservoirs is in local thermodynamic equilibrium with electrochemical potential $\mu_{\beta}$ and temperature $T_{\beta}$. The superconducting chemical potential $\mu_S$ sets the common Fermi energy, which we take from now on as the reference energy, $\mu_S=0$.
The passive subsystem is held at a lower temperature
($T\equiv T_N=T_S$) than that of the active subsystem ($T<T_H$).

The second term in Eq.~\eqref{eq_H} accounts for the
double quantum dot,
\begin{equation}\label{eq_dqd}
{\cal H}_{{\rm dqd}}=\underset{\alpha}{\sum}\varepsilon_{\alpha}\hat{n}_{\alpha}+U_{p}\hat{n}_{p,\uparrow}\hat{n}_{p,\downarrow}+U_{ap}\hat{n}_{a}\hat{n}_{p},
\end{equation}
where the number operators are defined as $\hat{n}_{p}=\sum_{\sigma}\hat{n}_{p,\sigma}=\sum_{\sigma}\hat{d}_{p,\sigma}^{\dagger}\hat{d}_{p,\sigma}$ for the passive dot,
and $\hat{n}_{a}=\hat{d}_{a}^{\dagger}\hat{d}_{a}$ for the active one. Here, $\hat{d}_{p,\sigma}$ and $\hat{d}_{a}$ denote  electron annihilation operators and $\varepsilon_{\alpha}$ are the dot energy levels. The interdot charging energy is $U_{ap}$, and $U_p$ is the (intradot) charging energy of the passive dot. 
%\sout{We neglect Coulomb intradot repulsion and spin effects if $\alpha=a$ because we hereafter focus on the currents generated in the passive dot. Thus, $U_p$ is nonzero in Eq.~\eqref{eq_dqd}.}
The coupling between the dots and the leads is given by the term:
%The active dot is tunnel coupled to a single (H) normal metallic reservoir and the passive dot is sandwiched between a normal (N) metal and a superconducting (S) electrode.
%Hence, the third term in Eq.~\eqref{eq_H} takes the form
\begin{gather}
\begin{aligned}
\label{eq_Ht}
{\cal H}_{t} &= \sum_{k,\sigma}\left(t_{N}\hat{d}_{p,\sigma}^{\dagger}\hat{c}_{k,N,\sigma}+t_{S}\hat{d}_{p,\sigma}^{\dagger}\hat{c}_{k,S,\sigma}+\rm{h.c.}\right)\\
&+ \sum_{k}\left(t_{H}\hat{d}_{a}^{\dagger}\hat{c}_{k,H}+\rm{h.c.}\right)
,
\end{aligned}
\end{gather}
where the $t_\beta$ are the dot-lead tunnel couplings. In the following, we consider the
wide-band approximation, in which case the tunnel hybridization strength is given by $\Gamma_{\beta}=2\pi|t_{\beta}|^{2}\rho_{0}^{\beta}$,
where $\rho_{0}^{\beta}$ is the corresponding electrode's density
of states in its normal state. 

\subsection{Power and efficiency}

Let $J_\beta$ be the heat flux and $I_\beta$ the charge current in terminal $\beta$. While charge conservation ensures $I\equiv I_N=-I_S$ and $I_H=0$, energy conservation in the conductor is only expressed as $J_N+J_S=P-J_H$, where \begin{equation}
\label{eq:power}
P=-VI
\end{equation}
is the power dissipated (Joule heating) by a charge flowing in favor of a voltage bias $V=(\mu_N-\mu_S)/e$. In that case, $P<0$.

In the absence of additional forces ($V=0$), the difference $T_H-T>0$ causes heat to be transferred from the active to the passive dot via the Coulomb coupling $U_{ap}$~\cite{hotspots}. A nonlocal thermoelectric engine is able to convert this injected heat flow into a charge current generated in the passive
subsystem. 
%generated by the conversion of $J_H$. 
Then, a finite voltage
$V$ can be applied to counteract $I$.
In the range of $V$ such that $P$ is positive, work can be done against an external load. The ratio
between the output power in the passive circuit
and the heat absorbed from the active one,
\begin{equation}\label{eq_eff}
    \eta = \frac{P}{J_H}
\end{equation}
is the thermodynamic efficiency of the engine, which is limited
by the Carnot bound, $\eta\le \eta_C=1-T/T_H$~\cite{benenti:2017}. For voltages larger than the stall voltage (a non-local analogue of the thermovoltage, where the thermoelectric current is compensated by electrons flowing in favor of the bias), power is dissipated and the engine stops working.
It is then necessary to calculate $P$, $J_H$ and $\eta$ to fully
describe the heat engine properties of our device. 

Sign changes of the heat currents define the other thermodynamic operations. When positive, $J_N$ and $J_S$ will characterise the cooling power of the system working as a refrigerator of the corresponding reservoir. A heat pump occurs when $J_H<0$.

\section{Andreev regime}
\label{sec:andreev}

\begin{figure}[t]
\includegraphics[width=\linewidth]{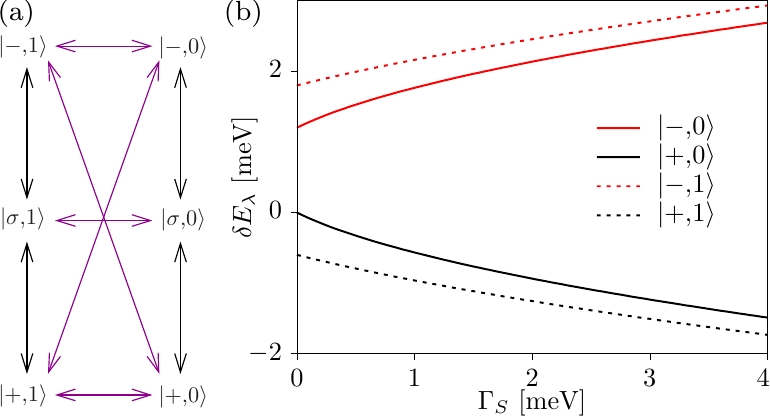}
\caption{\label{fig:andreevtransit}(a) Eigenstates of the proximitized quantum dot. The arrows represent the different transitions due to tunneling in the passive (black) and active systems (purple arrows). Note that the latest cannot change the parity of the passive system.
(b) Separation of the even and odd states, $\delta E_{\pm,n}{\equiv} E_{\pm,n}-E_{\sigma,n}$, as a function of the pairing, $\Gamma_S$, with $U_p=2U_{ap}=\unit[1.2]{meV}$ and $\varepsilon_a=0$.} 
\end{figure}

When the superconducting gap $\Delta$ is the largest energy scale of the problem, one can replace the superconductor and its coupling to the passive dot with a pairing term in Eq.~\eqref{eq_dqd}:
\begin{equation}
    {\cal H}_{\rm dqd} \to {\cal H}_{\rm dqd} + \Gamma_{S}(\hat{d}_{p,\uparrow}^{\dagger}\hat{d}_{p,\downarrow}^{\dagger}+{\rm {h.c.})},
\end{equation}
see e.g., Refs.~\cite{rozhkov:2000,Trocha:2014,Nigg:2015} for microscopic justifications of this approximation.
Then, ${\cal H}_{\rm dqd}$ can be exactly diagonalized
yielding the eigenenergies
\begin{align}
    \label{eq:Esig}
     E_{\sigma,n}&=\varepsilon_{p}+n(\varepsilon_{a}+U_{ap}) \\
    \label{eq:Epm}
     E_{\pm,n}&=n\varepsilon_{a}+A_{\mp,n},
\end{align}
with the associated eigenstates
\begin{align}
    \{\left|\sigma,n\right\rangle& \,,
    \left|\pm,n\right\rangle  =\mathcal{N}_{\pm,n}^{-1}(A_{\pm,n}\left|0,n\right\rangle -\Gamma_{S}\left|2,n\right\rangle )\},
    \label{eq:evenst}
\end{align}
represented in Fig.~\ref{fig:andreevtransit}(a). Here, $n=0,1$ keeps record of the charge number in the active dot and we have defined 
\begin{equation}
A_{\pm,n}=\tilde{\varepsilon}_{n}\pm\sqrt{\left(\tilde{\varepsilon}_{n}\right)^{2}+\Gamma_{S}^{2}},
\end{equation}
with $\tilde{\varepsilon}_{n}=\varepsilon_{p}+U_{p}/2+nU_{ap}$.
The states in the proximitized dot are expressed
in terms of the charge basis $\{0,\sigma,2\}$ and fall into
either the odd ($\left|\sigma\right\rangle$)
or the even ($\left|\pm\right\rangle$) charge sector.
The latter are important because they involve
coherent superpositions of states with 0 or 2 electrons,
and consequently contribute to the transfer of Cooper
pairs through the dot.

For small lead-dot couplings, a master equation
approach correctly describes the system dynamics in terms of sequential transitions between
the different states, as sketched in Fig.~\ref{fig:andreevtransit}(a). We get the stationary occupation of the different states, $P_\lambda$, by solving the set of equations
\begin{equation}
\label{eq:rateeq}
\sum_{\alpha\beta\kappa}\left(W_{\lambda\kappa}^{\alpha\beta}P_\kappa-W_{\kappa\lambda}^{\alpha\beta}P_\lambda\right)=0.
\end{equation}
The rates
$W_{\lambda\kappa}^{\alpha\beta}=\Gamma_{\lambda\kappa}^{\alpha\beta}+\gamma_{\lambda\kappa}^{\alpha\beta}$
for the transition $|\kappa\rangle\to|\lambda\rangle$
due to the tunneling of an electron from (to) lead $\beta$
in (out of) dot $\alpha$ are given respectively by
\begin{align}
    \Gamma_{\lambda\kappa}^{\alpha\beta}&=\Gamma_{\beta}|\langle\lambda|\hat{\delta}_{\alpha}^{\dagger}|\kappa\rangle|^{2}f_{\beta}\left(E_{\lambda}-E_{\kappa}\right)\\
    \gamma_{\lambda\kappa}^{\alpha\beta}&=\Gamma_{\beta}\bigl|\bigl\langle\lambda\bigl|\hat{\delta}_{\alpha}\bigr|\kappa\bigr\rangle\bigr|^{2}[1-f_{\beta}\left(E_{\kappa}-E_{\lambda}\right)],
\end{align}
where $f_\beta(E)=\{1+\exp[(E{-}\mu_\beta)/\kBT_\beta]\}^{-1}$
is the Fermi function. The operator $\hat\delta_\alpha$ annihilates an electron in dot $\alpha$ i.e., $\hat\delta_a=\hat d_a$ and $\hat \delta_p=\sum_\sigma\hat d_{p\sigma}$. These rates allow us to
determine both the charge current,
\begin{equation}\label{eq_I}
    I=e\underset{\lambda,\kappa}{\sum}\left(\gamma_{\lambda\kappa}^{p N}-\Gamma_{\lambda\kappa}^{p N}\right)P_\kappa,
\end{equation}
and the heat flux out of terminal $\beta$,
\begin{equation}\label{eq_J}
    J_{\beta} =\underset{\lambda,\kappa}{\sum}\left(E_{\lambda}-E_{\kappa}-\mu_\beta\right)\left(\gamma_{\kappa\lambda}^{\alpha \beta}P_\lambda-\Gamma_{\lambda\kappa}^{\alpha\beta}P_\kappa\right).
\end{equation}

\begin{figure}[t]
    \includegraphics[width=\linewidth]{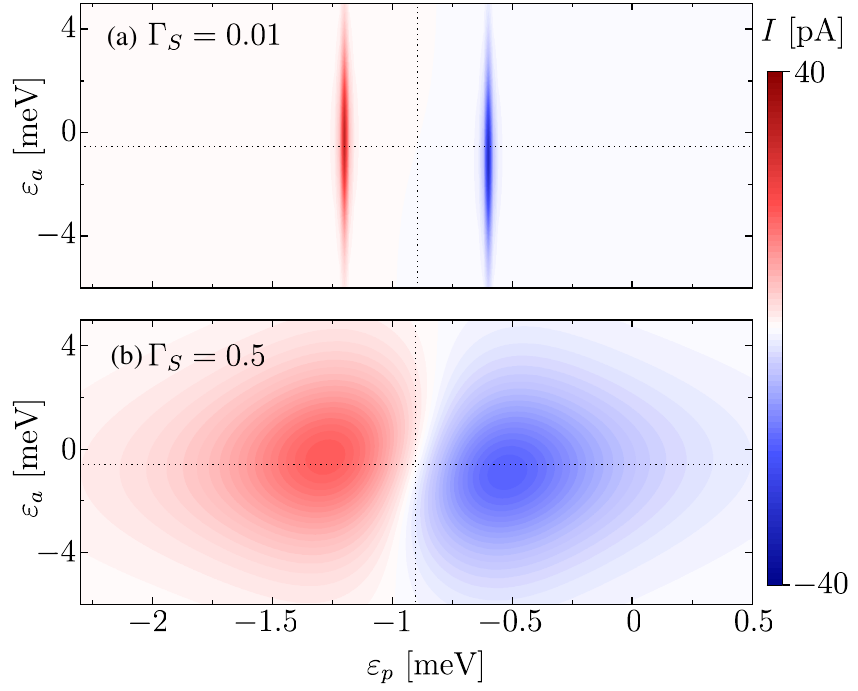}
    \caption{Charge current in the passive subsystem as a function of active and passive dot levels for (a)
    $\Gamma_S=\unit[0.01]{meV}$ %middle panel: $\Gamma_S=0.1$
    and (b) $\Gamma_S=\unit[0.5]{meV}$. Parameters (in meV): $k_B T_H=1.5$, $k_B T_C=1$, $\Gamma_H=\Gamma_N=0.01$,
    $U_{ap}=0.6$, $U_p=1.2$ and $V=0$. }
    \label{fig_curAndreev}
\end{figure}

The transitions between odd and even parity states can be both due to an electron or to a hole tunneling process. The relative rate at which one or the other process contributes more depends on the ratio $A_{\pm,n}/\Gamma_S$ that dictates the asymmetry of the even superpositions in Eq.~\eqref{eq:evenst}. It depends on the energy of the passive dot and, in particular, on the occupation of the active one, $n$.
In fact, the gap between even and odd
states is itself a function of $n$, as shown in Fig.~\ref{fig:andreevtransit}(b).
Hence, fluctuations of the charge in the active dot change the contribution of electron- or hole-like processes in the passive dot in a dynamical way. This results in a rectification effect that leads to a finite $I$ at $V=0$.  

%Let us first consider the unbiased case, $V=0$, in analogy with the drag effect. 
In Fig.~\ref{fig_curAndreev} we plot the thermoelectric current in the passive circuit as a function of the energies of the two dot levels, which can be shifted with external gate voltages~\cite{vanderwiel:2002}. 
%and resulting diagrams are easily accessible experimentally. \textcolor{blue}{I think it is not clear what we mean by diagrams here}
A finite current is generated in the passive circuit due to Coulomb interactions with the hotter active circuit.
%, in agreement with Ref.~\cite{mojtaba}.
Similarly to Ref.~\cite{mojtaba},
the passive system response occurs around the center of the stability diagram, where fluctuations of the charge of both dots are enhanced. However, while the drag current in Ref.~\cite{mojtaba} is driven by nonequilibrium fluctuations created by a dc bias applied across the active dot, the generated current here is driven by purely thermal means.

\begin{figure}
    \includegraphics[width=\linewidth]{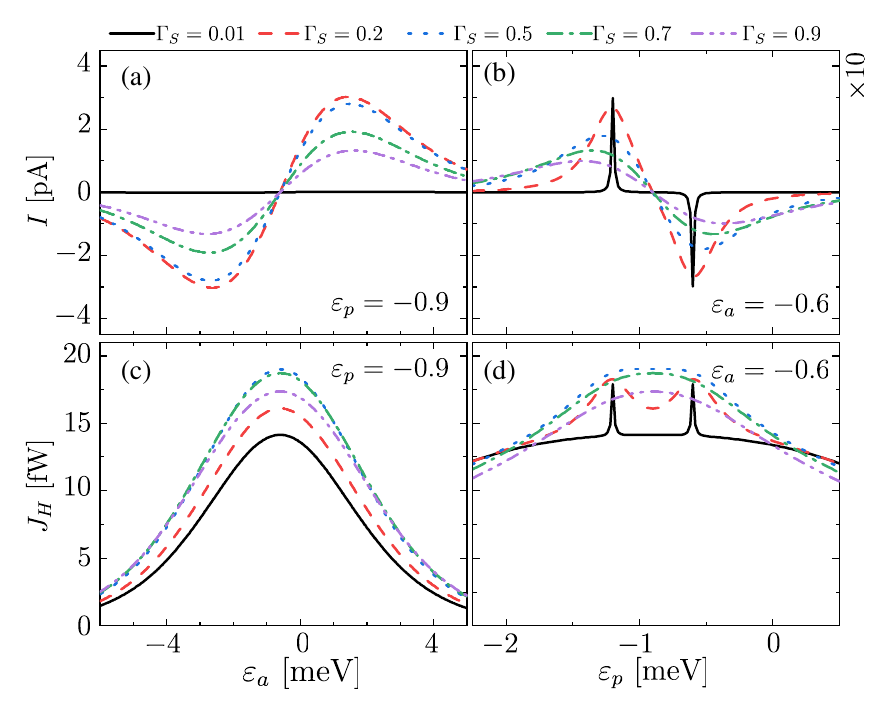}
    \caption{(a-b) Zero bias charge and (c-d) heat current from H for different couplings to the superconductor: $\Gamma_S=0.01$, $0.2$, $0.5$, $0.7$, and $\Gamma_S=\unit[0.9]{meV}$. The different panels show the dependence of the currents as a function of both the active (a), (c) and passive (b), (d) dot levels, along the dashed lines in Fig.~\ref{fig_curAndreev}(a) and (b). The current in panel (b) is divided by 10 to fit with the axis of panel (a). Other parameters are as in Fig.~\ref{fig_curAndreev}}
    \label{fig_heatAndreev}
\end{figure}

For small coupling to the S electrode,
the current changes sign by tuning $\varepsilon_p$, see Fig.~\ref{fig_curAndreev}(a). The current direction is determined by the character (electron- or hole-like)
of the dominant transitions, which is strongly dependent on the position of the passive dot energy.
The generated current can be seen at a much wider range of gate voltages when $\Gamma_S$ is further increased, see Fig.~\ref{fig_curAndreev}(b).
An important aspect is that the generated current changes sign in this case also when tuning the level position of the active dot, keeping $\varepsilon_p$ fixed, as can be seen in Fig.~\ref{fig_heatAndreev}(a). This way, both the magnitude and the sign of the passive current can be controlled by using the parameters of the active system as an external knob.
Linecuts of the currents along the dotted lines in Fig.~\ref{fig_curAndreev} are plotted in Fig.~\ref{fig_heatAndreev}(a) and (b) for clarity. 
%However, the current sign reversal phenomenon across the diagram boundaries is nicely preserved.
%\sout{For higher temperature differences, thermal smearing destroys this effect although the sign change is still present.}

Let us now investigate the heat currents.
To make a comparison, we show in Figs.~\ref{fig_heatAndreev}(c) and (d) the heat current $J_H=-J_N$  together with the charge current
through the passive dot. While $I$ has the characteristic
sign changes discussed above, $J_\beta$ does not change sign, as expected for heat between a hot and a cold reservoir in the absence of external work done on the system. 
%The heat flow for the active subsystem is positive because energy is extracted from the hot reservoir.
This energy is transferred through the
interdot repulsion $U_{ap}$ into the passive subsystem in the region where the charge states of the two quantum dots fluctuate strongly.
%Energy conservation is fulfilled in the form: $J_N+J_H=0$.
We recall that in the infinite gap approximation
the superconducting reservoir is only treated
as a source of pair correlations for the $p$ dot
and therefore we have two heat fluxes only.
In the next sections, we will relax this approximation
and study also the heat that flows in the superconductor.

Furthermore, the charge and heat currents in Fig.~\ref{fig_heatAndreev} show a nonmonotonic behaviour of the maximal charge current with the coupling to the superconductor: for small values of $\Gamma_S$, the generated current increases with the coupling to S, as expected. However, as it becomes comparable to and larger than the charging energy, both $I$ and $J_H$ get reduced. This is understood in terms of the increased splitting between the even and odd parity states, see Fig.~\ref{fig:andreevtransit}(b). For large $\Gamma_S$, the fluctuations between the lowest energy $|+,n\rangle$ state and the $|\sigma,n\rangle$ states (due to either the tunneling of an electron or a hole) are hence suppressed, as soon as $E_{\sigma,n}{-}E_{+,n}>\kBT$, in a similar way as the charging energy suppresses single-electron transport~\cite{kouwenhoven:1997}. These fluctuations are necessary both for the charge and the heat currents, resulting in a backaction on the thermal transport between H and N. Hence, pairing cooperates with $U_{ap}$ in blocking the current. Differently from Coulomb blockade, this pairing blockade effect is quantum coherent. 

\begin{figure}
    \centering
    \includegraphics[width=\linewidth]{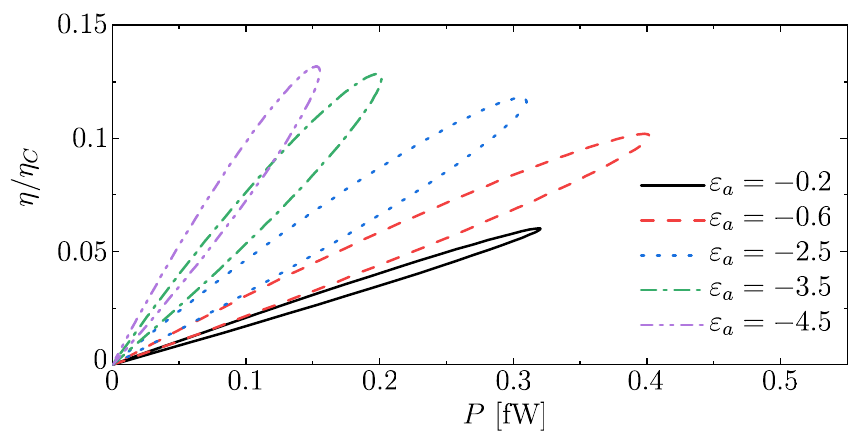}
    \caption{Thermoelectric performance of the heat engine in the Andreev regime for different values of $\varepsilon_a$. Power and efficiency are computed for voltages ranging between zero and the stall voltage, where $P=\eta=0$. Other parameters are as in Fig.~\ref{fig_curAndreev}(a).}
    \label{fig_effAndreev}
\end{figure}

We are now in a position to assess the thermoelectric efficiency $\eta$. Positive power $P$ is produced when the current flows against the applied voltage $V$. This occurs between $V=0$ and the stall voltage where the bias compensates the thermoelectric current (making $I=0$). 
%For positive $V$, this occurs when $I<0$. 
Thus, in our efficiency calculations we select the maximum negative current, which occurs around $\varepsilon_p=-(U_p+U_{ap})/2+U_{ap}/2$, as shown in Fig.~\ref{fig_heatAndreev}.
By tuning $V$, we calculate both $P$ and $\eta$, which both vanish at $V=0$ and at the corresponding stall voltage where $I=0$, see Eqs.~\eqref{eq:power} and \eqref{eq_eff}, as shown in Fig.~\ref{fig_effAndreev}.
%We then vary $V$ and consequently $P$, from which we can plot $\eta$ normalized to the Carnot limit as a function of $P$, as illustrated 
We observe that power and efficiency are optimized for different values of $\varepsilon_a$. When the gates are such that current is maximal (around $\varepsilon_a=\unit[-0.6]{meV}$), the power is enhanced. However, the efficiency increases when both $\varepsilon_a$ and $\varepsilon_a+U_{ap}$ go well below $\mu_H$, at the expense of reducing the fluctuations and hence the generated power.
In order to have a sizable performance, the value of $\Gamma_S$ needs to be optimized as well. We need it to be of the order of $A_{\pm,n}$ in order to maximize the effect of the fluctuations of the pairing wavefunction. 
%We find small values of the efficiency when $\Gamma_S$ is low since the generated current grows with increasing the coupling to the superconducting terminal. \MT{$\leftarrow$ Maybe it is better to say this in the caption of Fig.5?}
%Both the efficiency and the power increase as $\Gamma_S$ is increased. At some point, the effect is reversed, i.e., the efficiency decreases with further increases of $\Gamma_S$. 
In the limits where either  $\Gamma_S\gg\tilde{\varepsilon}_n$ or the opposite, the asymmetry between electron- and hole-like processes disappears and thereby the thermoelectric effect vanishes.

\section{Quasiparticle regime}
\label{sec:quasipart}

The other contribution to transport is due to quasiparticles. They are expected to dominate in different configurations: either when the passive dot level is close to the superconducting gap, when $\Gamma_S\rightarrow0$, or at high temperatures.
In these cases, tunneling of Cooper pairs
is strongly suppressed. In order to discriminate its contribution, we consider a simple model where the effect of the superconductor is only due to its density of states (DoS). The spin of the electron does not play a crucial role, so for simplicity we assume the limit $U_p\gg\kBT$ where a charge basis captures the main effect. %It is then sufficient to consider the charge states

We are then left with four states $|n_p, n_a\rangle$, with 0 or 1 electron in the passive and active dots.
%(infinite onsite interaction limit), although we keep the interdot interaction $U_{ap}$ finite.
The hopping rates in the passive subsystem are
%\begin{align}
%    \Gamma_{1n,0n}^{p,N}&=\Gamma_{N}f_{Nn},\\
%\Gamma_{1n,0n}^{p,S}&=\Gamma_{S}\nu_{n}f_{Sn},\\
%\gamma_{0n,1n}^{p,N}&=\Gamma_{N}(1-f_{Nn}),\\
%\gamma_{0n,1n}^{p,S}&=\Gamma_{S}\nu_{n}(1-f_{Sn}),
%\end{align}
%\begin{align}
%\Gamma_{1n,0n}^{p,\beta}&=\Gamma_{\beta}\nu_{\beta n}f_{\beta n},\\
%\gamma_{0n,1n}^{p,\beta}&=\Gamma_{\beta}\nu_{\beta n}(1-f_{\beta n}),
%\end{align}
\begin{align}
\Gamma_{1n,0n}^{p,\beta}{=}\Gamma_{\beta}\nu_{\beta n}f_{\beta n}\ \ \text{and}\ \ 
\gamma_{0n,1n}^{p,\beta}{=}\Gamma_{\beta}\nu_{\beta n}(1{-}f_{\beta n}),
\end{align}
where $\nu_{Nn}=1$, and $\nu_{Sn}$ is the normalized superconductor DoS.
Notably, we take into account the dependence on the active
dot occupation $n=0,1$ via the Coulomb repulsion strength
$U_{ap}$. This is better seen in the Fermi functions as
$f_{\beta n}=f_{\beta}(nU_{ap}+\varepsilon_{p})$
for $\beta$=N,S.
We model the superconductor DoS with the Dynes
form~\cite{dynes:1978},
\begin{equation}
    \nu_{Sn} = \left|
    {\Re}\,\frac{\varepsilon_p+nU_{ap}+i\chi}{\sqrt{(\varepsilon_p+nU_{ap}+i\chi)^2-\Delta^2}}
    \right|,
\end{equation}
where $\chi$ takes into account quasiparticle occupation in the gap.
The rates in the active dot read
\begin{align}
    \Gamma_{n1,n0}^{a,H}=\Gamma_{H}f_{H n}\quad\text{and}\quad
\gamma_{n0,n1}^{a,H}=\Gamma_{H}(1-f_{H n}),
\end{align}
where $f_{H n}=f_{H}(nU_{ap}+\varepsilon_{a})$. 
We can find the stationary occupation probabilities $P_{n_pn_a}$ by solving the system of equations~\eqref{eq:rateeq}. With these, we again calculate charge and heat currents using Eqs.~\eqref{eq_I} and~\eqref{eq_J}. 
%With all these, we can again calculate the charge
%\begin{equation}\label{eq_Iqp}
%    I_{N}^{\rm qp} =e\underset{n}{\sum}\left(\gamma_{0n,1n}^{p N}P_{1n}-\Gamma_{1n,0n}^{p N}P_{0n}\right),
%\end{equation}
%and heat currents
%\begin{gather}
%\label{eq_Jqp}
%\begin{aligned}
%    J_{\beta}^{\rm qp} &=\underset{n}{\sum}(\varepsilon_p{+}nU_{ap}{-}\mu_\beta)\left(\gamma_{0n,1n}^{p \beta}P_{1n}{-}\Gamma_{1n,0n}^{p \beta}P_{0n}\right)\\
%    J_{H}^{\rm qp} &=\underset{n}{\sum}(\varepsilon_a{+}nU_{ap}{-}\mu_H)\left(\gamma_{1n,0n}^{a l}P_{n1}{-}\Gamma_{0n,1n}^{a l}P_{n0}\right),
%\end{aligned}
%\end{gather}
%where $\beta=N,S$.

\begin{figure}
    \includegraphics[width=\linewidth]{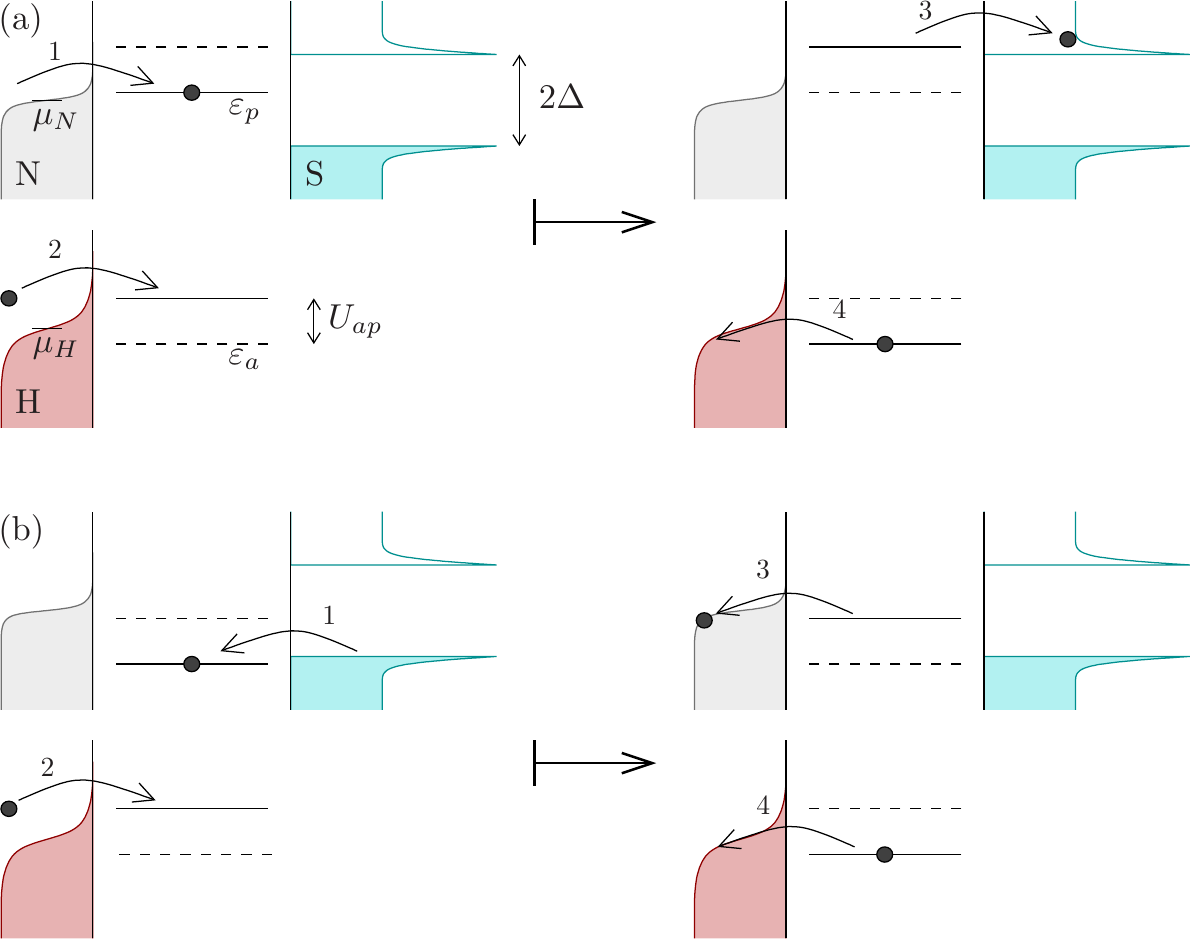}
    \caption{Quasiparticle transport enabling sequences at zero bias. A finite current is induced by a temperature difference $T_H>T_N=T_S$ whose sign depends on the passive dot energy. Charge flows (a) from $N$ to $S$ for   $\varepsilon_p-\mu_N\in(\Delta{-}U_{ap},\Delta)$, and (b) in the opposite direction when $\varepsilon_p-\mu_N\in(-\Delta{-}U_{ap},-\Delta)$. The order of the tunneling events is indicated.}
    \label{fig_procQP}
\end{figure}

\begin{figure}
    \includegraphics[width=\linewidth]{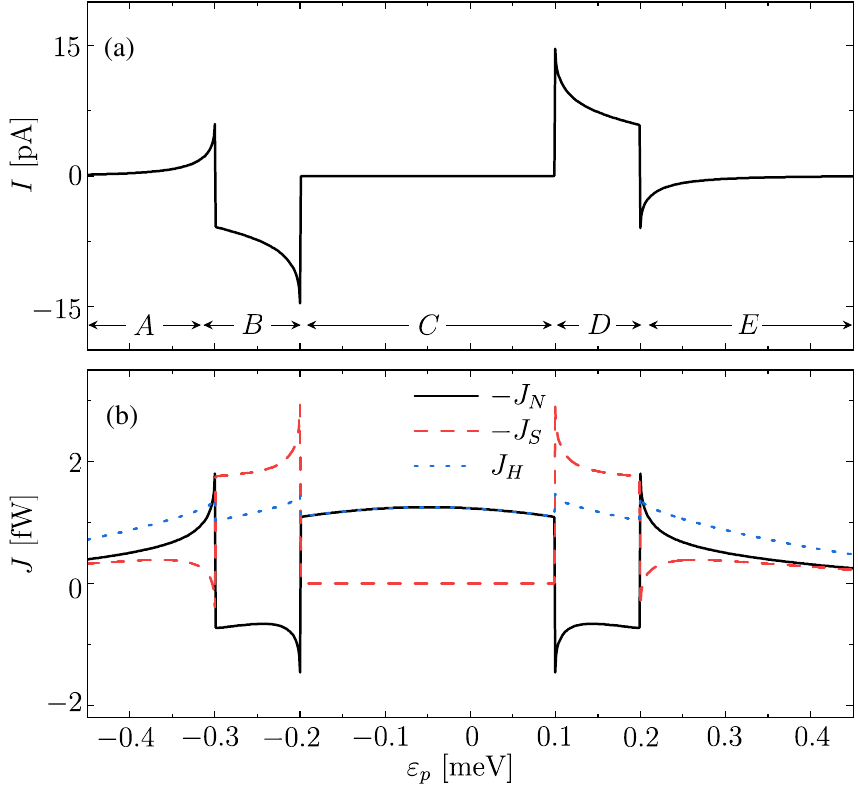}
    \caption{Zero bias (a) charge and (b) heat currents in the different terminals in the quasiparticle regime. They are plotted as functions of the passive dot energy. 
    Parameters: $\Delta=\unit[0.2]{meV}$, $k_B T_H=\unit[0.3]{meV} $, $k_B T=\unit[0.2]{meV}$,
    $U_{ap}=\unit[0.1]{meV}$, $\varepsilon_a=-U_{ap}/2$, $\Gamma_H=\Gamma_N=\Gamma_S=\unit[0.01]{meV}$, and $\chi=10^{-6}$.
    }
    \label{fig_heatQP}
\end{figure}

It was shown in Ref.~\cite{hotspots} (and later confirmed experimentally~\cite{thierschmann:2015}) that a similar configuration (based there on all-normal reservoirs) leads to a non-local thermoelectric response provided the tunneling rates in the passive system are energy-dependent and left-right asymmetric. In our hybrid configuration, this property is enabled by the superconducting DoS: tunneling is suppressed at energies $\varepsilon_p+nU_{ap}$ within the gap, and enhanced at energies close to the coherence peaks, as sketched in Fig.~\ref{fig_procQP} for two different level positions. It can be shown that in the case where the tunneling couplings $\Gamma_\beta$ are energy independent, a zero-bias current \begin{equation}
\label{eq:IqpV0}
I(V{=}0)\propto(\nu_{S0}-\nu_{S1})(f_{N0}-f_{N1})
\end{equation}
is generated~\cite{mojtaba}. 
%The BCS DoS hence restricts electron tunneling to occur close to the gap edges, a property that has been exploited recently for power sources~\cite{Marin-Suarez:2022,pekola:2022}. 
This current is shown in Fig.~\ref{fig_heatQP}(a). As expected, $I=0$ when both energies $\varepsilon_p+nU_{ap}$ lie in the gap [region with $\varepsilon_p-\mu_N\in(-\Delta,\Delta{-}U_{ap})$, marked as C].
The conditions for transport appear in regions of width given by the Coulomb repulsion $U_{ap}$, marked as B [with $\varepsilon_p-\mu_N\in(-\Delta{-}U_{ap},-\Delta)$] and D [with $\varepsilon_p-\mu_N\in(\Delta{-}U_{ap},\Delta)$] in Fig.~\ref{fig_heatQP}(a). There, electrons/holes around the Fermi energy in N (such that $\nu_{S0}=0$ at those energies) can exchange energy with the passive dot and be transferred over/below the gap (when $\nu_{S1}\neq0$). In this configuration, the charge current is antisymmetric around $\varepsilon_p=-U_{ap}/2$. Note that the change of sign around the particle-hole symmetry point is opposite to the one obtained in the Andreev regime discussed in Sec.~\ref{sec:andreev}. Also differently from that case, the charge current does not change sign with the position of the active dot, due to the lack of coherence in the passive dot.

\begin{figure}[t]
    \includegraphics[width=.9\linewidth]{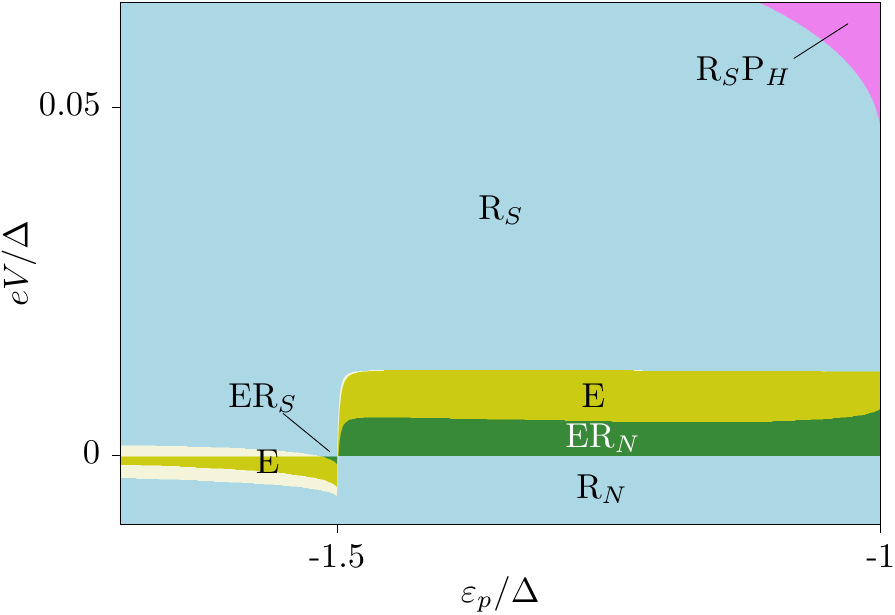}
    \caption{Map of the different thermal operations as a function of the passive dot energy and the applied voltage. Different colors mark different thermodynamic operations, as indicated (white means no useful operation). They are labeled with E, when the system works as a heat engine, and with R$_\alpha$ and P$_\alpha$, when it works as a refrigerator or a heat pump in terminal $\alpha$. Parameters: $\kBT=\Delta$, $\kBT_H=1.1\Delta$, $\varepsilon_a=-0.3\Delta$, $\chi=10^{-4}$. Other parameters as in Fig.~\ref{fig_heatQP}. 
    %Same parameters as in Fig.~\ref{fig_curAndreev}, with $\Gamma_S=0.5$. 
    }
    \label{fig_multitask}
\end{figure}

The same transport windows are evident in the heat transport, as shown in Fig.~\ref{fig_heatQP}(b). In region C, where transport in terminal S is avoided by the gap, all the heat injected from H is absorbed by the normal contact. Out of this region, the heat current sign gives insights of the relevant mechanism for the generated current. In region B electrons below the gap in S enter the passive dot when the active one is empty and tunnel out below the chemical potential of N once the active dot has been occupied. This way, $I<0$, the superconductor is heated up ($J_S<0$) and the normal contact is refrigerated ($J_N>0$), as sketched in Fig.~\ref{fig_procQP}(b). In the opposite region D, electrons tunnel from over $\mu_N$ into the passive dot, with an empty active dot, and tunnel out over the superconducting gap after gaining an energy $U_{ap}$ from the interaction with an electron having occupied the passive dot, see Fig.~\ref{fig_procQP}(a). In this case, while the charge flows in the opposite direction, $I>0$, one still finds $J_S<0$ and $J_N>0$, as expected. Noticeably, we find that close to $\varepsilon_p-\mu_N\approx\pm(\Delta+U_{ap})$ (in the frontiers of A-B and D-E regions), the peaks of the superconducting DOS enable an additional change of sign in the charge current that cools the superconductor. In all these cases (except for region C), the system works as a hybrid thermal engine~\cite{manzano:2020} capable of simultaneously generate power and cool (either terminal N or S) from a single resource (heat in reservoir H). This possibility is further explored in Fig.~\ref{fig_multitask} in the presence of a finite voltage. This way, we find an additional hybrid operation where high enough voltages are able to reverse the heat current $J_H$, while $J_S>0$ (violet region in Fig.~\ref{fig_multitask}, labelled as R$_S$P$_H$).  Hence, the system works both as a refrigerator of S and a heat pump into the hottest reservoir, H. Tuning the voltage facilitates the cooling of the N or S terminals (light blue regions in Fig.~\ref{fig_multitask}), very much like a usual quantum dot Peltier refrigerator~\cite{edwards:1993,prance:2009}. Here these operations are affected by the coupling to the other dot. The configuration map in Fig.~\ref{fig_multitask} is repeated by inverting over the $V=0$ and $\varepsilon_p-\mu_N=-U_{ap}/2$.

\begin{figure}[t]
    \includegraphics[width=\linewidth]{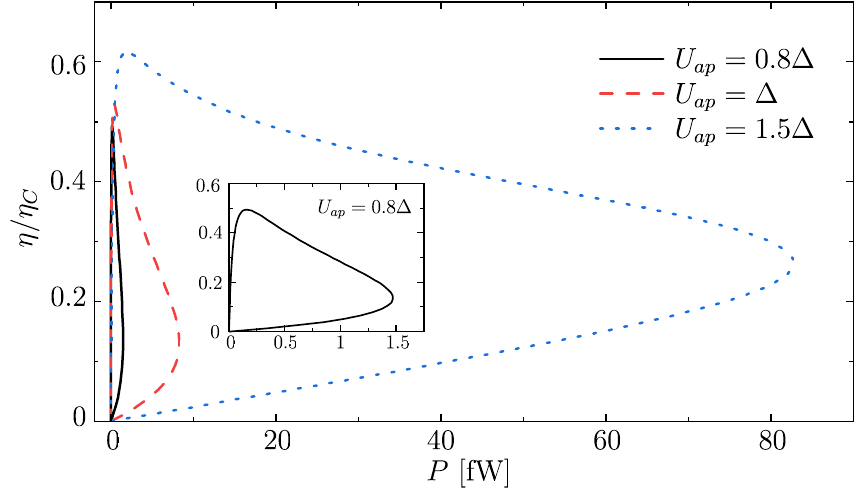}
    \caption{Thermoelectric efficiency in the quasiparticle regime, for different interaction $U_{ap}$. Each point is optimized with respect to the value of $\varepsilon_p$. The inset zooms the curve for $U_{ap}=0.8\Delta$ in, for clarity. Other parameters as in Fig.~\ref{fig_heatQP}.
    }
    \label{fig_effPQP}
\end{figure}

The asymmetric filtering of some electronic tunneling events, here facilitated by the superconducting DoS, is expected to enable optimal heat to power conversion in terms of power and efficiency~\cite{hotspots}. 
Consider for instance region D, and the scheme in Fig.~\ref{fig_procQP}(a): the gap avoids the electron tunneling from N to subsequently tunnel to S without having exchanged an energy $U_{ap}$ with the active system. Once this is done, the coherence peak makes it more favorable to tunnel to S than back to N. This way, most of the cycles for which an amount of energy $U_{ap}$ is transferred from the active to the passive system result in an electron being transferred from N to S.
%On one hand, the gap suppresses $\nu_{S0}$ when $-\Delta<\varepsilon_p<\mu_N$. On the other hand, if $\varepsilon_p+U_{ap}\ge \Delta$, $\nu_{S1}$ is enhanced thanks to the BCS peak. 
%This way, the current in Eq.~\eqref{eq:IqpV0} is maximized. The same applies for the case $\varepsilon_p+U_{ap}\le -\Delta$ and $\mu_N<\varepsilon_p+U_{ap}<\Delta$, exchanging the subindices $0\leftrightarrow1$. 
This is confirmed in Fig.~\ref{fig_effPQP}, where power increases by two orders of magnitude with respect to the Andreev regime, with the efficiency being doubled. Furthermore, for low $T$, high efficiencies approaching $\eta_C$ are obtained at low power (but still one order of magnitude larger) by taking advantage of the BCS peak. The maximal efficiency is hence limited by the sharpness of the coherence peak, which is affected by the Dynes parameter, $\chi$. In this case, the thermoelectric performance improves when increasing $U_{ap}$, such that electrons gain enough energy from H to overcome the gap. Increasing temperature smears the effect of the spectral features and hence affects the performance.

\section{Intermediate regime}
\label{sec:keldysh}

In this section we explore the thermoelectric performance of the engine in the intermediate regime where the temperature is low and the superconducting energy gap $\Delta$ is comparable to other energy scales. Then it is important to take into account the contributions from both Andreev and quasiparticle transport mechanisms in the calculations. This is possible by employing the non-equilibrium Green's functions (NEGF) technique~\cite{mojtaba}, which furthermore takes the finite linewidth of the quantum dot states (which is neglected in the sequential tunneling master equations discussed in previous sections) into account. 

Within this formalism, the different currents through terminal $\beta$=N,S,H, can be calculated by using
\begin{align}
I_{\beta}&=  \frac{e}{2\hbar}\int\frac{d\omega}{2\pi}\text{Tr}\left(\hat\sigma_z\cal{I}_{\beta}\right)\\
%\end{align}
%for charge, and
%\begin{align}
J_{\beta}&= \frac{1}{2h}\int\frac{d\omega}{2\pi}\text{Tr}\left[\left(\omega I_{2}-eV_{\beta}\hat{\sigma}_{z}\right){\cal I}_{\beta}\right],
%\text{Tr}\left[\hat{\sigma}_{z}\times\right.\nonumber \\
% & (G_{\alpha}^{R}\left(\omega\right)\Sigma_{\alpha,\beta}^{<}\left(\omega\right)+G_{\alpha}^{<}\left(\omega\right)\Sigma_{\alpha,\beta}^{A}\left(\omega\right)\nonumber \\
% & \left.-\Sigma_{\alpha,\beta}^{R}\left(\omega\right)G_{\alpha}^{<}\left(\omega\right)-\Sigma_{\alpha,\beta}^{<}\left(\omega\right)G_{\alpha}^{A}\left(\omega\right)\right],
\end{align}
for charge and heat, respectively, with
\begin{gather}
\label{eq:currop}
\begin{aligned}
\cal{I}_{\beta}= ~ &%\text{Tr}\left[
G_{\alpha}^{R}\left(\omega\right)\Sigma_{\alpha,\beta}^{<}\left(\omega\right)+G_{\alpha}^{<}\left(\omega\right)\Sigma_{\alpha,\beta}^{A}\left(\omega\right)%\right.
 \\
 & %\left.
 -\Sigma_{\alpha,\beta}^{R}\left(\omega\right)G_{\alpha}^{<}\left(\omega\right)
 -\Sigma_{\alpha,\beta}^{<}\left(\omega\right)G_{\alpha}^{A}\left(\omega\right).%\right].
\end{aligned}
\end{gather}
Here the terminal index $\beta$ fixes the index $\alpha=p,a$ of the quantum dot coupled to it, and
$I_2$ and $\hat\sigma_i$ are the $2{\times}2$ identity and Pauli matrices, respectively. The retarded and lesser selfenergies are given by $\Sigma_{a,H}^{R}\left(\omega\right)=-i\Gamma_{a,H}$, $\Sigma_{p,N}^{R}\left(\omega\right)=-i\Gamma_{p,N}I_{2},$ and
$\Sigma_{p,S}^{R}\left(\omega\right)=-i\Gamma_{S}b\left(\omega\right)(I_{2}-\Delta\omega^{-1}\hat\sigma_{x}),$ where
$b\left(\omega\right)\equiv\left|\omega\right|(\omega^{2}-\Delta^{2})^{-1/2}\theta\left(\left|\omega\right|-\Delta\right)-i\omega(\Delta^{2}-\omega^{2})^{-1/2}\theta\left(\Delta-\left|\omega\right|\right)$, with
the Heaviside step function, $\theta\left(E\right)$, and $\Sigma_{\alpha,\beta}^<(\omega)=-2i{\rm Im}(\Sigma_{\alpha,\beta}^R)f_\beta(\omega)$. The interacting retarded and lesser components of the Keldysh Green's function of dot $\alpha$, $G_{\alpha}^{R/<}$, are calculated self-consistently by using the Dyson and the Keldysh equations
\begin{align}
\label{eq:Gar_interacting}
G_{\alpha}^{R}&=\left\{ \left[g_{\alpha}^{R}\right]^{-1}-\Sigma_{\alpha,{\rm {int}}}^{R}\right\} ^{-1}\\%\end{equation}
%as well as the Keldysh equation 
%\begin{equation}
G_{\alpha}^{<}&=G_{\alpha}^{R}\left(\Sigma_{\alpha,{\rm {leads}}}^{<}+\Sigma_{\alpha,{\rm {int}}}^{<}\right)G_{\alpha}^{A},\label{eq:Glpinteracting}
\end{align}
where $g_{\alpha}^{R}$ is the non-interacting retarded Green's function of dot $\alpha$, and $\Sigma_{\alpha,{\rm {int}}}$ is the corresponding interaction selfenergy of the dot $\alpha$ due to its Coulomb interaction with other charges in the dots. We refer the interested reader to the supplementary material in Ref.~\cite{mojtaba} for the explicit expressions of these self energies and Green's functions and also for the details of the numerical procedure used to solve the above equations selfconsistently. Here we chose the parameters such that the system falls in the intermediate regime where both Andreev and quasiparticle mechanisms contribute. In particular, their relative contribution is controlled by the coupling $\Gamma_S$.

\begin{figure}
    \includegraphics[width=\linewidth]{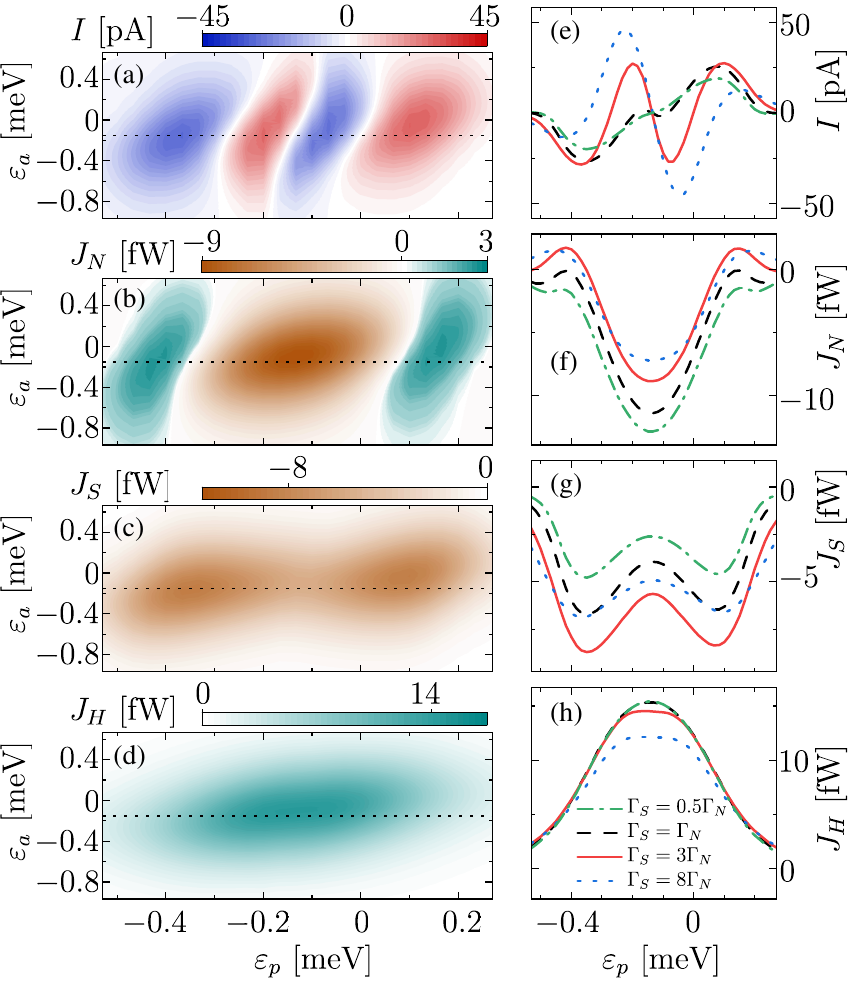}
    \caption{(a)-(d) Charge $I$ and heat currents $J_N$, $J_S$ and $J_H$ as functions of the active and passive dot energies for $\Gamma_S=3\Gamma_N$. (e)-(h) Line cuts of left panels at $\varepsilon_a=-(U_{ap}+U_p)/2=\unit[-0.15]{meV}$ (indicated by dashed lines in left panels) and for different values of $\Gamma_S=0.5\Gamma_N$, $\Gamma_N$, $3\Gamma_N$ and $8\Gamma_N$. Other parameters are $\Delta=\unit[0.2]{meV}$, $\kBT_H=2\kBT=\unit[0.2]{meV}$, $\Gamma_H=\Gamma_N=\unit[0.01]{meV}$, $U_p=1.6U_{ap}=\unit[0.16]{meV}$ and $V=0$.
    }
    \label{fig_NEGF-EaEp}
\end{figure}

As we learned in the previous sections, the contribution of the two different mechanisms (Cooper pair and quasiparticle dominated transport) depends strongly on the position of the quantum dot levels. The Andreev (pairing) induced current is finite around the particle-hole symmetric point, see Fig.~\ref{fig_curAndreev}. Differently, the quasiparticle contribution occurs close to the gap borders, see Fig.~\ref{fig_heatQP}. Hence, their competition results in a rich behaviour of the current generated at $V=0$. This is shown in Fig.~\ref{fig_NEGF-EaEp}(a): in the center of the stability diagram the Andreev contribution dominates, with the current sign being as predicted by the rate equation description in Sec.~\ref{sec:andreev}. As the passive dot level deviates from this region, the quasiparticle contribution starts to dominate and the current changes sign, in agreement with the results in Sec.~\ref{sec:quasipart}.
Importantly, for fixed values of $\varepsilon_p$, the charge current changes sign as a function of $\varepsilon_a$. This effect was found in the Andreev regime, cf. Figs.~\ref{fig_curAndreev}(b) and \ref{fig_heatAndreev}(a), and is hence a signature of quantum coherence persisting in the intermediate regime. This is interesting since it allows us to control the generated current by only using the active dot parameters: $T_H$ generates the current and $\varepsilon_a$ controls its direction.

The different contributions are also visible in the heat currents, plotted in Figs.~\ref{fig_NEGF-EaEp}(b)-(d). As expected, the heat current injected from H is always positive (for $V=0$). It is restricted to the region where charge fluctuations in both dots are present~\cite{koski:2015}, see Fig.~\ref{fig_NEGF-EaEp}(d), emphasizing the role of electron-electron interaction in the heat exchange between the systems. The latter occurs as well in the passive system currents, with two particular aspects to point out: heat always flows into the superconductor ($J_S<0$) but is slightly suppressed by the gap for $\varepsilon_p$ close to the particle-hole symmetry point. Most interestingly, $J_N$ changes sign at the crossover between the Andreev and the quasiparticle regimes, leading to cooling, see Fig.~\ref{fig_NEGF-EaEp}(b). Hence, we recover the ER$_N$ operation shown in Fig.~\ref{fig_multitask} as soon as a small voltage $V$ is applied.

The relative contribution of the two mechanisms depends strongly on the parameters. Crucially,  it depends on the coupling to the superconductor, because the Andreev reflection is suppressed for low $\Gamma_S$. This is shown in Figs.~\ref{fig_NEGF-EaEp}(e)-(h), where cuts of the left-column panels at $\varepsilon_a=-(U_{ap}+U_p)/2$ (marked by dotted lines) are plotted for different $\Gamma_S$. Remarkably, Fig.~\ref{fig_NEGF-EaEp}(e) shows how the Andreev contribution to the charge current dominates for large $\Gamma_S$ and totally disappears for sufficiently small couplings. Note that the quasiparticle contribution gets correspondingly reduced but can still be visible for large $\Gamma_S$ by the effect of the gap border. In this case, the sharp features in Fig.~\ref{fig_heatQP} due to the gap are smeared here due to the finite level-width of the passive dot. We observe a different behaviour in Fig.~\ref{fig_NEGF-EaEp}(f), where the contribution of the strong gap features to the cooling of N are favored by increasing the coupling to the superconductor.  
%\MT{This is mainly because for larger values of $\Gamma_S/\Gamma_N$, the strong coupling to the normal terminal softens the effect of the nonlinear density of states of the $S$ terminal around gap border. }
While $J_S$ gets strongly reduced for low $\Gamma_S$, cf. Fig.~\ref{fig_NEGF-EaEp}(g), the injected current from the hot bath is barely unaffected, except for large couplings, see Fig.~\ref{fig_NEGF-EaEp}(h).
The latter can be attributed to the pairing blockade effect introduced in Sec.~\ref{sec:andreev}. Note however a stronger nonmonotonic behaviour of $J_S$ in Fig.~\ref{fig_NEGF-EaEp}(g), that is caused by the increasing role  at high $\Gamma_S$ of pair tunneling, which does not carry heat. 
%The latter can be attributed to the splitting of the passive dot states, see Eq.~\eqref{eq:Epm}]: as $\Gamma_S$ increases and becomes larger than the charging energy, the occupation of the $|+,n\rangle$ states will block the dynamics. On top of the charging energy, charge fluctuations in the passive system will hence require an additional energy due to pairing (a kind of coherent enhancement of Coulomb blockade of heat transport). %This will hence reduce

\begin{figure}[t]
    \includegraphics[width=\linewidth]{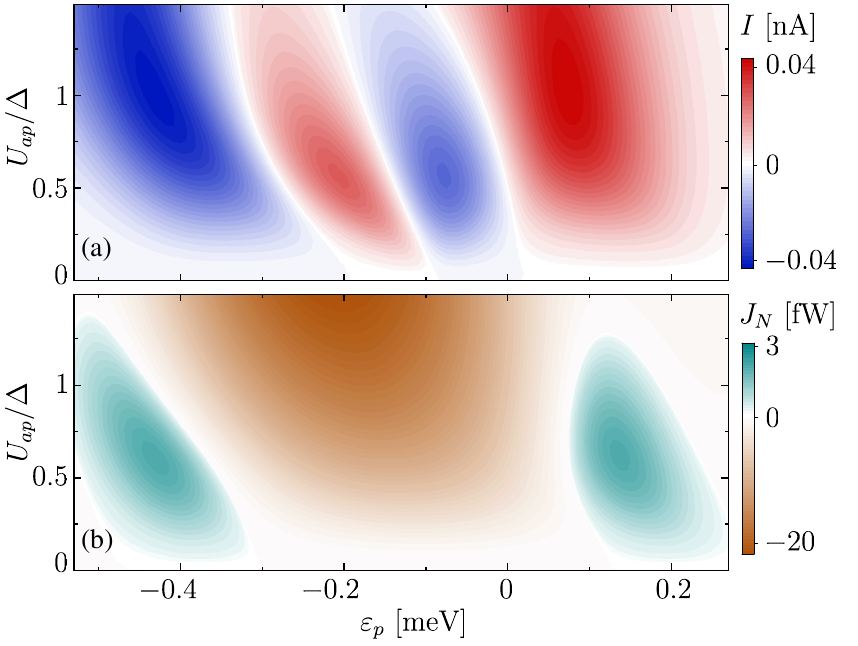}
    \caption{(a) Charge current, $I$, and (b) heat current, $J_N$, as functions of the interdot Coulomb interaction strength $U_{ap}$ and the passive dot energy for $\Gamma_S=3\Gamma_N$. $\varepsilon_a=\unit[-0.15]{meV}$ and other parameters are as in Fig.~\ref{fig_NEGF-EaEp}.   }
    \label{fig_NEGF-UapEp}
\end{figure}

The interplay of the two processes furthermore depends on the interaction between the two quantum dots, $U_{ap}$. This is shown in Fig.~\ref{fig_NEGF-UapEp}. The quasiparticle contribution to the charge current is maximal when $U_{ap}\lesssim\Delta$ i.e., when quasiparticles gain/lose enough energy from the interaction with the active dot to overcome the gap, see Fig.~\ref{fig_NEGF-UapEp}(a). For the same reason, the cooling of terminal N, due to quasiparticles, is suppressed for $U_{ap}>\Delta$, see Fig.~\ref{fig_NEGF-UapEp}(b). The Andreev contribution is on the contrary reduced by the interaction, as it reduces the hybridization of the even parity states (essential for the current generation via pairing) when it becomes large compared to $\Gamma_S$, see Eq.~\eqref{eq:evenst}. For large $U_{ap}$, the Andreev contribution is washed out by quasiparticles, as expected. 

\begin{figure}[t]
    \includegraphics[width=\linewidth]{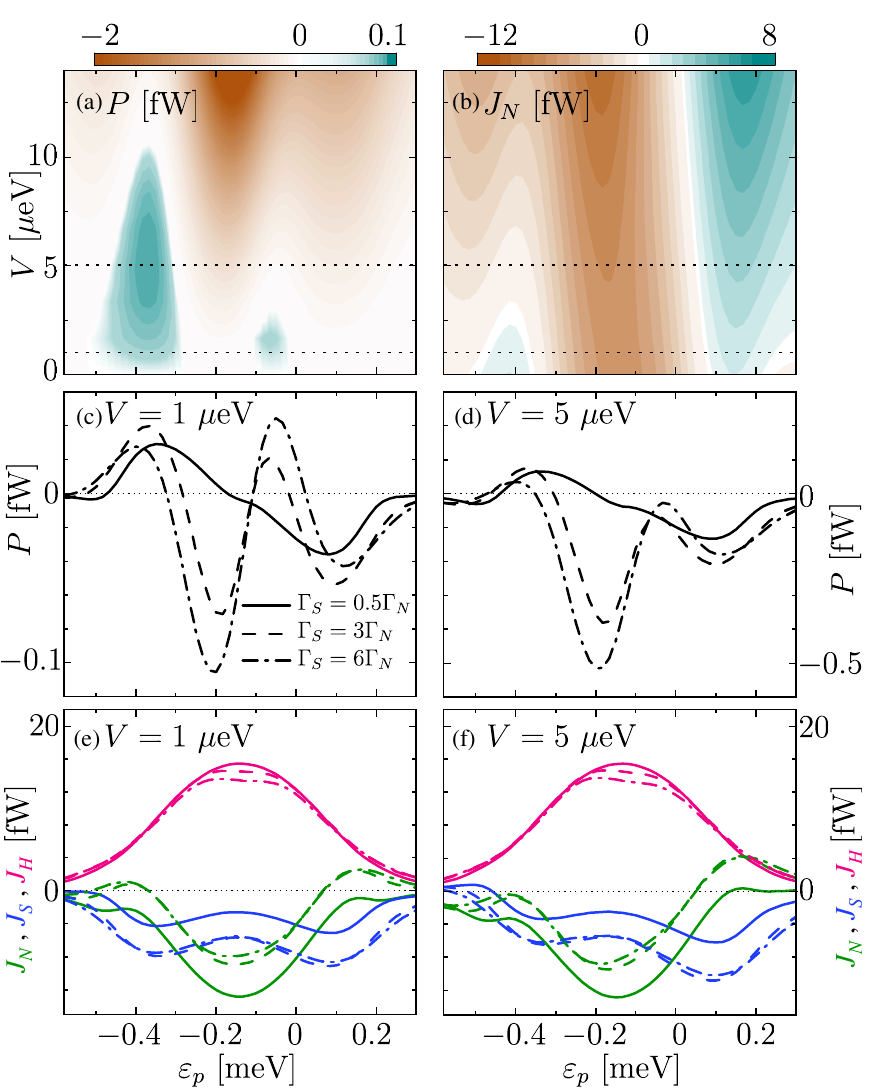}
    \caption{(a) Power and (b) normal terminal heat current in the low voltage regime as functions of voltage and the passive dot energy. Panels (c)-(e) show line-cuts of the power and the different heat currents for different $\Gamma_S$ and two different voltages indicated by dotted lines in (a) and (b). Same parameters as in Fig.~\ref{fig_NEGF-EaEp}, with $\varepsilon_a=\unit[-0.15]{meV}$.}
    \label{fig_NEGF-power-VbEp}
\end{figure}

The generated current leads to useful power in the presence of a finite bias voltage. This affects the properties of the heat currents as well. In Fig.~\ref{fig_NEGF-power-VbEp} we show the generated power and the heat current in N. Figures~\ref{fig_NEGF-power-VbEp}(a) and (b) show the low voltage behaviour, with line-cuts for fixed $V$ plotted in Figs.~\ref{fig_NEGF-power-VbEp}(c)-(f). In the low (but positive) voltage regime, finite power is generated in two regions as we tune $\varepsilon_p$, which correspond to the quasiparticle and Andreev dominated regimes. The quasiparticle dominated region (around $\varepsilon_p\approx\unit[-0.4]{meV}$) results in larger power and larger stall voltages for this intermediate configuration. The cooling of terminal N discussed above coexists with the power production, though the limiting voltage for having $J_N>0$ is smaller than the stall voltage for $P>0$, compare Figs.~\ref{fig_NEGF-power-VbEp}(a) and (b). This is consistent with the behaviour shown in Fig.~\ref{fig_multitask}, where the E and ER$_N$ are adjacent for quasiparticle induced transport. For the configuration chosen here, the efficiency is reduced to $\eta\approx0.02\eta_C$, which we attribute to the finite width of the passive level and to the fact that the competition of the two mechanisms unavoidably reduces the generated power, without necessarily reducing the transferred heat $J_H$. 

The line-cuts in Figs.~\ref{fig_NEGF-power-VbEp}(c)-(f) show $P$ and $J_\beta$, for different $\Gamma_S$. The same behaviour that we observed for the current is shown here for the power: as we decrease the coupling to S, the quasiparticle contribution dominates and give $P>0$ even close to the particle-hole symmetric point. The injected current from H is little affected. However, we observe that in the region with Cooper pair contributions ($\varepsilon_p$ around the particle-hole symmetry point), $J_H$ decreases with $\Gamma_S$, while it increases as quasiparticles take over (for larger $|\varepsilon_p|$), confirming the coherent nature of the reduction of $J_H$ in the gap, in agreement with the pairing backaction effect identified in Sec.~\ref{sec:andreev}. Note however that here we are in the regime $\kBT>\Gamma_S$.

Decreasing $\Gamma_S$ suppresses the ER$_N$ regions, as seen by comparing Figs.~\ref{fig_NEGF-power-VbEp}(c) and (e) for $\varepsilon_p\approx\unit[-0.4]{meV}$, as expected for a quasiparticle-only effect. For even more negative $\varepsilon_p$, we find cooling of the superconductor induced by $V$, for small $\Gamma_S$, see Figs.~\ref{fig_NEGF-power-VbEp}(d) and (f). However, the operation ER$_S$ is not found, which we attribute to the effect of the finite linewidth of the quantum dot states. We speculate that $P>0$ and $J_S>0$ may coexist for even smaller $\Gamma_N$ and $\Gamma_S$. Unfortunately this will also considerably reduce the heat flows.

\begin{figure}
    \includegraphics[width=\linewidth]{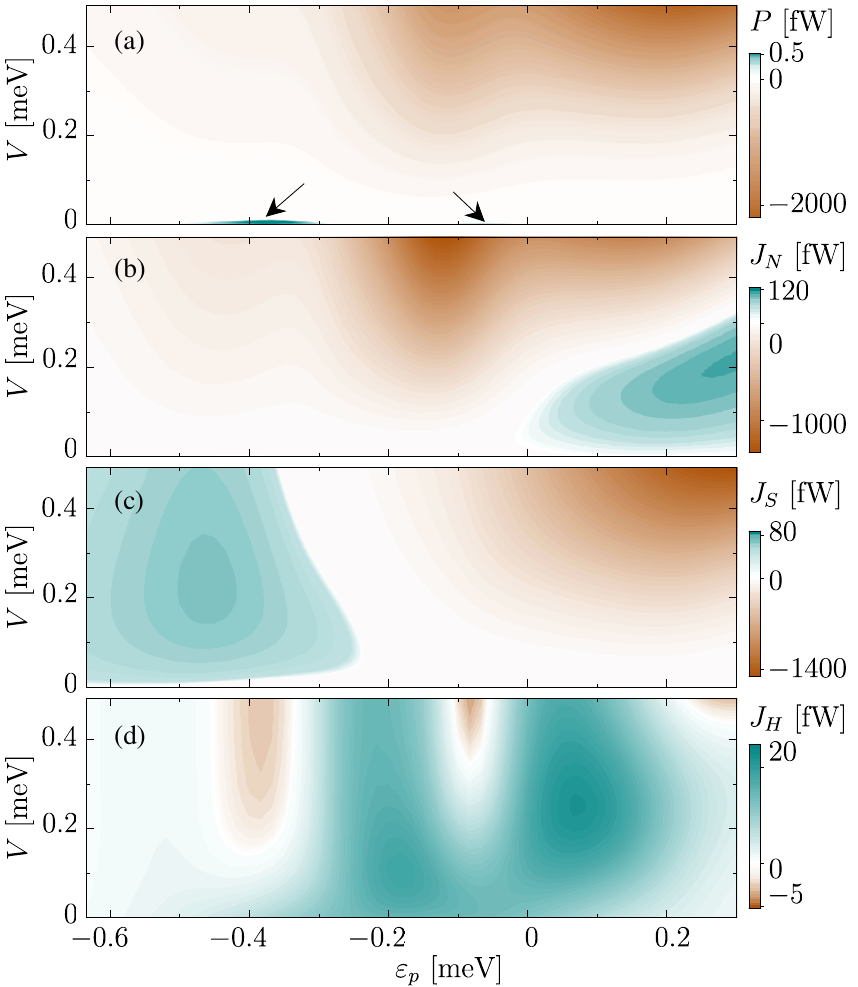}
    \caption{Plots of (a) power, (b) $J_N$, (c) $J_S$ and (d) $J_H$, as functions of the bias voltage $V$ and the passive dot energy $\varepsilon_p$. Arrows in panel (a) indicate the regions where $P>0$, shown in Fig.~\ref{fig_NEGF-power-VbEp}(a) for clarity. The same configuration as in Fig.~\ref{fig_NEGF-power-VbEp} is considered, for a larger range of $V$. Note that the regions of $P>0$ are almost invisible in panel (a) for being restricted to very low voltages.
    }
    \label{fig_NEGF-VbEp}
\end{figure}

The currents in the presence of larger voltages are shown in Fig.~\ref{fig_NEGF-VbEp}. There, the charge current flowing in favor of the bias leads to power dissipation, $P<0$. The arrows in Fig.~\ref{fig_NEGF-VbEp}(a) mark the regions where power is produced, as zoomed in in Fig.~\ref{fig_NEGF-power-VbEp}(a). Dissipation leads to heating terminal N, see Fig.~\ref{fig_NEGF-power-VbEp}(b), except for large and positive $\varepsilon_p$, where electrons tunnel from N into the passive dot over $\mu_N$. Voltage also allows for the heat extracted from the superconductor to increase, see Fig.~\ref{fig_NEGF-VbEp}(c), again due to quasiparticles (we remind that Cooper pairs do not lead to heat flows, as discussed in Sec.~\ref{sec:andreev}). Note the cooling power $J_S$ is maximal for voltages of the order of the gap. This process is related to refrigerators based on NIS tunnel junctions~\cite{giazotto:2006}, here mediated by the fluctuations in a quantum dot, see also Ref.~\cite{sanchez_correlation_2017}. Additionally, the dissipated Joule power reverts the heat flow at terminal H close to the gap borders, see Fig.~\ref{fig_NEGF-VbEp}(d) for $\varepsilon_p\approx\unit[-0.4]{meV}$  and $\varepsilon_p\gtrsim\unit[0.2]{meV}$, which makes the system work as a heat pump and recovers the R$_S$P$_H$ operation discussed in Sec.~\ref{sec:quasipart}. Remarkably, for high enough $V$, we observe a subgap region (around $\varepsilon_p\approx\unit[-0.1]{meV}$) where $J_H<0$, i.e., induced by Andreev processes and their correlation with the charge fluctuations of the active dot.  

\section{Conclusions}
\label{sec:conclusions}

We have shown how a hybrid conductor composed by a quantum dot coupled to a normal and a superconductor contact can work as a quantum heat engine, when  capacitively coupled to a second quantum dot in contact with a hot terminal. The injection of heat is mediated by the interaction of electrons in different dots, whose charge fluctuates. The asymmetry of the normal-superconductor contact induces a zero-bias current via two competing mechanisms: on one hand, the superconducting density of states acts as an efficient energy filter for quasiparticle tunneling. On the other hand, the hybridization of even parity states due to pairing depends on the occupation of the hot dot, hence changing the relative contribution of empty and doubly occupied states to the superposition and affecting the electron- or holelike character of the electrons tunnelling from N. This way, the symmetry of a wavefunction is used as the piston of an autonomous quantum heat engine.  

We have analyzed the limiting cases where these two contributions operate separately by using simple models described by quantum master equations. More realistic configurations where both effects coexist are treated numerically with NEGF techniques. The relative contribution of the two mechanisms can be controlled with gate voltages acting on the quantum dot levels. The contribution of Andreev processes dominates for states close to the chemical potential of the superconductor, while quasiparticle processes are enhanced for transitions via states aligned with the gap. The opposite contributions of the two mechanisms lead to changes of sign in the generated zero-bias current as the passive dot level is tuned which are not present in all-normal configurations~\cite{dare:2017,walldorf:2017}. The quantum coherent character of this mechanism manifests in an additional change of sign of the generated current as the active dot level is tuned. In this way, the coupled quantum dot acts not only as the heat source but also as an external knob of the response. We also identify a backaction of pairing on the active system in the form of a coherent pairing-Coulomb blockade of the injected heat current. 
%\sout{Differently, the quasiparticle transport occurs for states close to the gap edges.} 
The gap and the coherence peak in the superconducting DoS are beneficial for the conversion of heat into power, which reaches high efficiencies in the quasiparticle regime. Furthermore, it allows the system to work as a hybrid engine where several thermodynamic operations (power generation, cooling and/or heat pumping) coexist. Most of these features survive in the intermediate regime, where the finite linewidth of the quantum dot states introduces a limitation to the performance of the device. The robustness of the Andreev induced currents in this regime makes this system appealing for the experimental realization of quantum themodynamic engines.

\acknowledgements
We acknowledge P. Burset and G. Steffensen for useful comments. Work funded by Iran Science Elites Federation (ISEF) and by Spanish State Research Agency through Grants Nos.\ PID2019-110125GB-I00, PID2020-117347GB-I00, PID2020-117671GB-I00 and RYC2016-20778 and through the Severo Ochoa and María de Maeztu Program for Centers and Units of Excellence in R\&D (Nos.\ MDM2017-0711 and CEX2018-000805-M) funded by MCIN/AEI/10.13039/501100011033.

%\begin{thebibliography}{}
%\bibliographystyle{apsrev}

%%%%%%%%%%%%%%%%%%%%%%%%%%%%%%%%%%%%%%%%%%%%%%%%%%%%%%%%%%%%%%%%
\bibliography{biblio.bib}
%%%%%%%%%%%%%%%%%%%%%%%%%%%%%%%%%%%%%%%%%%%%%%%%%%%%%%%%%%%%%%%%

\end{document}